\documentclass[pdflatex,sn-mathphys-ay,oneside]{sn-jnl}

\usepackage{graphicx}
\usepackage{amsmath,amssymb,amsfonts}
\usepackage{booktabs}
\usepackage{multirow}
\usepackage{hyperref}
\usepackage{xcolor}
\usepackage{float}
\usepackage{enumitem}

\newcommand{\bra}[1]{\langle #1 |}
\newcommand{\ket}[1]{| #1 \rangle}

\begin{document}

\title[Quantum Machine Learning for Few-Shot Impersonation Detection in Digital Account Opening]{\textbf{Quantum Machine Learning for Few-Shot Impersonation Detection in Digital Account Opening}}

\author*[1]{\fnm{Javiera} \sur{Canales}}\email{javiera.canales@coredevx.com}
\author[2]{\fnm{Jose} \sur{Ortega}}
\author[1]{\fnm{Maximiliano} \sur{Araya}}

\affil[1]{\orgdiv{CoreDevX LABTAV  -  Advanced Technologies Laboratory}}
\affil[2]{\orgname{Banco de Cr\'editos e Inversiones (BCI)}, \orgaddress{\city{Santiago}, \country{Chile}}}

\abstract{Identity impersonation fraud may emerge only after digital account opening, requiring an early post-onboarding approach based on behavioral and transactional signals observed shortly after activation. Confirmed cases remain rare, creating extreme class imbalance and material operational risk for digital banks. We evaluate quantum machine learning (QML) against classical one-class methods in this setting using a real 2025 cohort of 3,419 digital account openings from Banco de Cr\'editos e Inversiones (BCI), Chile, including 8 confirmed impersonation-fraud cases (0.23\% prevalence). The data were provided through an industry-research collaboration between BCI and CoreDevX LABTAV under operational constraints. The best-performing configuration used only 2 qubits and was executed on CoreDevX's SpinQ Triangulum II, a 3-qubit nuclear magnetic resonance (NMR) device, enabling hardware validation beyond classical simulation. The quantum one-class classifier based on Automatic Quantum Feature Mapping (AQFM) achieves 88\% recall (7 of 8 frauds detected) and 9.21\% precision, corresponding to approximately 70\% fewer false positives than the best-performing classical baseline at comparable recall. When its 2025-calibrated decision threshold is applied unchanged to the independent 2026 cohort, AQFM detects 20 of 31 frauds (64.5\% recall) with 8.81\% precision and remains the best-performing model among those evaluated.}


\maketitle

\section{Introduction}
\label{sec:introduction}

Identity impersonation associated with digital account opening is an escalating concern for banking platforms. Unlike application fraud that is detected and blocked during onboarding, the cases examined in this study concern accounts that successfully complete the opening process and are subsequently used for fraudulent activity. We therefore adopt an \emph{early post-onboarding approach}: account-opening information is combined with behavioral and transactional signals observed shortly after activation to identify impersonation fraud. Although the fraud manifests after the account has been opened, confirmed cases remain extremely rare, often fewer than ten across thousands of accounts. In this regime, supervised machine learning models struggle to learn discriminative boundaries, and financial institutions often rely on manual review processes that do not scale with digital growth in customer acquisition \cite{chandola2009anomaly, abdallah2016fraud_survey}.

The operational implications are substantial. Each undetected fraudulent account may result in regulatory exposure, reputational damage, and direct financial loss. Conversely, incorrectly flagging legitimate applications introduces friction into the onboarding process and consumes analyst resources. Under extreme label scarcity, the precision-recall trade-off becomes acute: high recall typically entails a large number of false positives, limiting the efficiency of automated screening systems. This risk is likely to intensify as generative AI tools lower the barrier to crafting convincing synthetic identities, potentially increasing both the volume and sophistication of impersonation attacks on digital onboarding channels \cite{europol2023deepfake}.

This extreme data scarcity places fraud screening in the regime of one-class or anomaly-detection methods \cite{khan2014occ_survey}. Classical one-class techniques - such as One-Class SVM \cite{scholkopf2001ocsvm}, Isolation Forest \cite{liu2008isolation}, and distance-based approaches \cite{ramaswamy2000knn} - can provide reasonable recall but typically incur high false-positive rates when the positive class contains fewer than ten examples, and standard imbalance remedies (SMOTE \cite{chawla2002smote}, cost-sensitive learning \cite{elkan2001foundations}, threshold adjustment \cite{fernandez2018learning}) lose effectiveness in this regime (detailed in Section~\ref{sec:background}).

The limitations of classical one-class methods motivate an evaluation of quantum machine learning (QML) under the same data constraints. Theory suggests that some quantum models may have favorable generalization properties in low-data settings \cite{huang2021power}, while quantum feature maps can represent decision boundaries that are difficult to obtain with linear models in the original feature space \cite{havlicek2019supervised}. Evidence from operational banking data, however, remains limited, particularly for one-class models designed for noisy intermediate-scale quantum (NISQ) hardware \cite{preskill2018quantum}.

This study was conducted as a joint industry-research collaboration between CoreDevX LABTAV (Advanced Technologies Laboratory) and Banco de Cr\'editos e Inversiones (BCI), spearheaded by BCI's Corporate Innovation Division. It evaluates quantum one-class classifiers using a real 2025 cohort that links completed digital account openings with early post-onboarding activity, where only 8 out of 3,419 accounts were confirmed as impersonation fraud. The main contributions are as follows:

\begin{enumerate}[itemsep=0pt,topsep=0pt]
    \item We demonstrate that, on the 2025 cohort, a quantum model achieves 88\% recall (7 of 8 frauds) with 9.21\% precision, reducing false positives by approximately 70\% compared to the best classical one-class baseline at comparable recall. When the decision threshold calibrated on 2025 training data is applied unchanged to the independent 2026 cohort, AQFM detects 20 of 31 frauds (64.5\% recall) with 8.81\% precision, retaining the strongest overall performance among the evaluated quantum and classical models.
\item We show that, among the quantum configurations evaluated, the 2-qubit circuit consistently outperforms alternatives with up to 23 qubits on this dataset. This reduced circuit size enabled validation on a real quantum computer  -  the SpinQ Triangulum II, a 3-qubit NMR device  -  beyond classical simulation.
\end{enumerate}

The remainder of the paper is organized as follows. Section~\ref{sec:background} situates the problem within the relevant literature. Section~\ref{sec:setting} describes the dataset, evaluation protocol, and modeling approach. Section~\ref{sec:evaluation} presents the empirical results. Section~\ref{sec:implications} discusses practical implications and limitations. Section~\ref{sec:conclusion} concludes.

\section{From Fraud Screening to Extreme Few-Shot Learning}
\label{sec:background}

\subsection{Classical Fraud Detection Under Extreme Imbalance}

Fraud detection systems in banking typically combine rule engines with supervised machine learning \cite{abdallah2016fraud_survey}. When fraud labels are scarce, the problem shifts toward one-class classification, where a model is trained exclusively on normal examples and deployed to identify deviations \cite{khan2014occ_survey}. Classical one-class methods - including OC-SVM \cite{scholkopf2001ocsvm}, Isolation Forest \cite{liu2008isolation}, distance-to-centroid approaches, and statistical Z-score filters \cite{hodge2004survey} - have been widely surveyed \cite{chandola2009anomaly}.

In extreme low-prevalence settings, these methods often provide reasonable recall but suffer from low precision. The operational consequence is that fraud analysts must manually review many false-positive alerts for each true fraud detected, undermining the efficiency gains that automated screening is meant to provide \cite{west2016intelligent}. Traditional remedies for class imbalance, such as SMOTE \cite{chawla2002smote}, cost-sensitive learning \cite{elkan2001foundations}, and threshold adjustment \cite{fernandez2018learning}, are challenged when the minority class contains fewer than ten examples, because synthetic resampling cannot reliably approximate the true fraud distribution.

\subsection{Quantum Machine Learning Models}

Three families of quantum classifiers are relevant to this study.

\textbf{Variational quantum classifiers (VQCs)} \cite{schuld2020circuit, schuld2019quantum} encode classical data into quantum states via parameterized feature maps and classify points based on measurement outcomes. A VQC consists of a fixed data-encoding stage followed by a parameterized ansatz whose weights are optimized classically to minimize a loss function. VQCs are the most general building block of near-term quantum machine learning \cite{cerezo2021variational}.

\textbf{Quantum support vector machines (QSVMs)} \cite{havlicek2019supervised, rebentrost2014quantum} replace the classical kernel with a quantum kernel that evaluates inner products between data points mapped into a high-dimensional Hilbert space. The quantum feature map can separate patterns that are not linearly separable in classical space, and the classifier is obtained from a standard SVM trained on the resulting quantum kernel matrix.

\textbf{Automatic Quantum Feature Mapping (AQFM)} \cite{altares2021automatic} uses a genetic algorithm to design quantum feature maps for QSVMs. Each candidate circuit is represented by a bit string that specifies gates and rotation angles. Crossover, mutation, and recombination generate new candidates, and selection retains circuits that improve a predefined objective. Unlike a VQC or a conventional QSVM with a fixed architecture, AQFM searches over circuit structures. In this study, the resulting feature map is detached from the SVM stage and evaluated directly: the expectation value of the circuit output provides the one-class anomaly score (Appendix~\ref{app:math}).

Published quantum anomaly-detection studies have examined density estimation \cite{liu2018quantum_anomaly} and variational circuits \cite{herr2021anomaly}, mainly on synthetic or benchmark data. Their performance on regulated financial data has received little empirical study.

\subsection{Gaps in Quantum One-Class Classification and Financial Imbalance}

Most work on quantum kernels and variational classifiers considers supervised learning with examples from each class \cite{havlicek2019supervised, schuld2020circuit, cerezo2021variational}. Far less attention has been given to the one-class case, in which training data contain normal observations but few or no labeled anomalies. This distinction matters here because the model must learn the normal-data boundary without using confirmed fraud cases during training.

Extreme imbalance occurs in account-opening fraud, money laundering, market manipulation, and rare credit defaults, where confirmed cases may be scarce relative to the number of transactions or accounts. One-class methods are commonly used in such settings, but their false-positive volume can make review queues impractical. Financial applications of quantum computing have concentrated on portfolio optimization, risk analysis, and derivative pricing \cite{herman2023portfolio, orus2019quantum_finance, egger2020quantum_finance, stamatopoulos2020option}. The present study instead examines a quantum one-class classifier on banking data under operational constraints.

\section{Operational Setting and Modeling Approach}
\label{sec:setting}

\subsection{Banking Dataset and Impersonation Characteristics}

The 2025 dataset was provided by Banco de Cr\'editos e Inversiones (BCI) and spans 18 months of digital account openings. After removing identifiers and non-predictive features, the final dataset contains 3,419 records with 21 engineered features, of which 8 are confirmed account-opening impersonations (0.23\% prevalence). The features used in this study encompass account-opening information, customer characteristics, and transactional behavior observed after the account is opened. All features were normalized to $[0,1]$ to accommodate angle encoding in quantum circuits. Numerical missing values were imputed using median strategies. Categorical variables were one-hot encoded, and redundant categories were consolidated.

\subsection{Evaluation Under Extreme Data Scarcity}

Given the extreme class imbalance, we adopt a one-class classification paradigm: models are trained exclusively on normal (non-fraudulent) accounts and then applied to an evaluation set containing all 8 fraud cases plus normal records. This protocol mirrors operational deployment, where new fraud patterns are unknown at training time.

We report deterministic metrics: recall, precision, and F1-score. Recall measures the fraction of confirmed frauds detected - critical because each undetected fraudulent account carries regulatory, reputational, and financial risk. Precision measures the proportion of alerts that correspond to true fraud - i.e., the screening efficiency, since each false alert consumes analyst review time. F1-score provides a harmonic summary of both. To characterize model performance across operating thresholds, we additionally report AUC-PR (Area Under the Precision-Recall curve), which summarizes the precision-recall trade-off across all decision thresholds and is particularly informative under extreme class imbalance. The business objective of the screening system is to maximize both recall and precision simultaneously: high recall ensures frauds are not missed, while high precision keeps the manual review burden operationally sustainable.

\subsection{Quantum Modeling Strategy}

Classical data cannot be directly processed by a quantum circuit: it must first be normalized and then encoded into quantum states. The normalization step maps numerical features to a bounded range compatible with the encoding scheme, while the encoding step associates each classical feature vector with a quantum state on which the circuit operates. This classical-to-quantum interface is a defining component of any quantum machine learning model and is commonly referred to as the \emph{feature map} \cite{schuld2019quantum, havlicek2019supervised}.

Amplitude, basis, and angle encoding offer different trade-offs among expressivity, qubit count, and circuit depth. We use angle encoding: each rescaled feature $x_i$ controls a single-qubit rotation. This encoding was considered for QSVM, VQC, and AQFM. Amplitude encoding is more compact in qubit count but requires a deeper state-preparation circuit.

The supervised QSVM implementation considered here requires labeled examples from both classes to define its separating boundary, whereas the training set contains no confirmed fraud cases. It would also require an $N \times N$ quantum kernel matrix, resulting in millions of circuit evaluations for this dataset. VQC and AQFM require one forward pass per observation at inference. We therefore excluded QSVM from the full evaluation and retained VQC and AQFM.

\subsection{Quantum One-Class Classification with QML}

The quantum models operate under a one-class classification paradigm: trained exclusively on normal account data, the model learns a representation of the legitimate distribution; at inference time, the anomaly score indicates deviation from this learned manifold. The decision threshold is calibrated on the training anomaly score distribution, and all datapoints above the threshold are considered anomalies or, in this case, potential frauds.

\subsection{Simulation and Deployment Infrastructure}

All quantum model training and simulation were conducted using NVIDIA CUDA-Q with GPU-accelerated statevector simulation (RTX Blackwell architecture). Hybrid classical optimization was performed entirely in simulation. The trained circuit was subsequently exported, translated into spinQit, and executed on a SpinQ Triangulum~II 3-qubit NMR quantum processor ($T_1 \sim 6$\,s, $T_2 \sim 300$\,ms). Hardware was used exclusively for inference validation; no parameter updates were performed on the quantum processor. Classical baselines (Isolation Forest, KNN anomaly detection, Distance Centroid, Z-Score) were implemented in Python with standard configurations. Detailed hyperparameters are listed in Appendix~\ref{app:experiments}.

\section{Real Data Evaluation}
\label{sec:evaluation}

\subsection{Classical Screening Baselines}

Table~\ref{tab:classical} reports the performance of five classical one-class methods, evaluated over the full 2025 cohort of 3,419 records (3,411 normal, 8 fraud) at the 88\% recall operating point (7 of 8 frauds detected), alongside AUC-PR values summarizing each model's precision-recall behavior across all thresholds.

At this operating point, classical models exhibit uniformly low precision. KNN Anomaly - the best-performing classical model at this threshold - achieves 3.0\% precision with approximately 226 false positives across 3,411 normal records. OC-SVM (RBF) achieves 1.43\% precision with roughly 482 false positives. Other models generate between 405 and 695 false positives. These results illustrate the intrinsic difficulty of anomaly detection when only 8 fraudulent cases are embedded within 3,411 legitimate accounts.

\begin{table}[h]
\centering
\caption{Performance of classical one-class methods on the full 2025 cohort of 3,419 records. AUC values are AUC-PR (Area Under the Precision-Recall curve). Precision is reported at 88\% recall (7 of 8 frauds). F1 is reported at 100\% recall (all 8 frauds flagged).}
\label{tab:classical}
\begin{tabular}{@{}lccc@{}}
\toprule
\textbf{Method} & \textbf{AUC-PR} & \textbf{Prec@7/8} & \textbf{F1} \\
\midrule
OC-SVM (RBF)           & 0.078 & 1.43\%   & 0.022 \\
KNN Anomaly            & 0.022 & 3.00\%   & 0.048 \\
Isolation Forest       & 0.031 & 1.00\%   & 0.015 \\
Distance Centroid      & 0.016 & 1.70\%   & 0.027 \\
Z-Score Anomaly        & 0.015 & 1.21\%   & 0.024 \\
\bottomrule
\end{tabular}
\end{table}

\subsection{Quantum Model Performance}

Section~\ref{sec:setting} explains why QSVM was excluded and VQC and AQFM were retained. Table~\ref{tab:quantum} reports their selected results. At 88\% recall (7 of 8 frauds detected), AQFM reaches 9.21\% precision with 69 false positives. At the same recall, KNN, the strongest classical baseline at this operating point, reaches 3.0\% precision with 226 false positives.

\begin{table}[h]
\centering
\caption{Performance of the AQFM quantum classifier on the 2025 cohort. The AQFM decision threshold was set at the 97.9th percentile of the training anomaly-score distribution ($\tau=0.0681816637$); scores equal to the threshold were classified as anomalies ($s\geq\tau$).}
\label{tab:quantum}
\begin{tabular}{@{}lcccc@{}}
\toprule
\textbf{Model} & \textbf{Recall} & \textbf{Precision} & \textbf{F1} & \textbf{FP} \\
\midrule
AQFM (2 qubits)  & 88\%  & 9.21\%  & 0.167 & 69 \\
VQC (Ry - Rx - Rz)  & 63\%  & 2.1\%  & 0.041 & 233 \\
\bottomrule
\end{tabular}
\end{table}

The AQFM model achieves this result as a quantum one-class classifier. At 88\% recall, classical KNN generates 226 false alerts and therefore requires 233 total reviews to identify 7 confirmed frauds, whereas AQFM generates 69 false alerts and requires 76 total reviews. This corresponds to a reduction of approximately 70\% in false-alert volume at identical recall.

\subsection{Threshold Transfer to a 2026 Cohort}

To assess temporal generalizability, the full pipeline was re-evaluated on an independent 2026 cohort of 5,143 digital account openings containing 31 confirmed impersonation cases (0.60\% prevalence), provided under the same collaboration. For every model, the decision threshold was calibrated on its 2025 training-score distribution at the same quantile (97.9th percentile) as the AQFM calibration of Table~\ref{tab:quantum}, and applied to the 2026 cohort without any re-tuning, so no information from the 2026 labels was used to set the operating points. The 2026 export omits one 2025 feature (Var\_2) and adds one numeric feature (Var\_47); preprocessing is otherwise identical.

\begin{table}[h]
\centering
\caption{Performance on the 2026 cohort (5,143 records, 31 frauds) with decision thresholds calibrated on 2025 training data (97.9th percentile of training scores) and applied unchanged. Thresholds are expressed in each model's own score scale and are not comparable across models.}
\label{tab:transfer2026}
\begin{tabular}{@{}lcccc@{}}
\toprule
\textbf{Model} & \textbf{Recall} & \textbf{Precision} & \textbf{FP} & \textbf{AUC-PR} \\
\midrule
AQFM (2 qubits)    & 64.5\% & 8.81\% & 207 & 0.191 \\
OC-SVM (RBF)       & 48.4\% & 5.08\% & 280 & 0.085 \\
Isolation Forest   & 22.6\% & 7.61\% & 85  & 0.051 \\
KNN Anomaly        & 0\%    & 0.00\% & 75  & 0.055 \\
Distance Centroid  & 0\%    & 0.00\% & 75  & 0.033 \\
Z-Score Anomaly    & 0\%    & 0.00\% & 75  & 0.033 \\
\bottomrule
\end{tabular}
\end{table}

The AQFM threshold is the only one that transfers while maintaining strong detection: it flags 20 of 31 frauds (64.5\% recall) at 8.81\% precision with 207 false positives. OC-SVM (RBF) and Isolation Forest transfer to lower-recall operating points (48.4\% and 22.6\% recall, respectively), while the distance-based methods (KNN Anomaly, Distance Centroid, Z-Score Anomaly) detect no frauds at their transferred thresholds, reflecting the outlier-dominated tails of their training-score distributions. Across all thresholds, the AUC-PR of AQFM (0.191) is more than twice that of the best classical baseline (OC-SVM RBF, 0.085).

\subsection{Confidence Intervals and Statistical Considerations}

Exact binomial confidence intervals (Clopper-Pearson) were calculated separately for each temporal cohort. For the 2025 cohort, the 95\% confidence interval for recall (7 of 8 frauds detected) is $[47.3\%, 99.7\%]$, and the interval for precision (7 of 76 predicted positives) is $[3.78\%, 18.06\%]$. On the independent 2026 cohort, AQFM detected 20 of 31 frauds, yielding 64.5\% recall with a 95\% confidence interval of $[45.4\%, 80.8\%]$, and 8.81\% precision (20 of 227 predicted positives) with an interval of $[5.46\%, 13.28\%]$. The intervals are not pooled because the cohorts represent distinct time periods and the 2026 analysis is intended as a temporal transfer test. Their width reflects the limited number of confirmed frauds rather than an artifact of model performance. The calculations are provided in Appendix~\ref{app:experiments}.

\subsection{Model Capacity and Qubit Scaling}

The relationship between qubit count and performance in this study does not follow a monotonic pattern. We compared AQFM configurations ranging from 2 to 13 qubits and several configurations of VQC with 23 qubits. The extended results are provided in Appendix~\ref{app:experiments}; the pattern is summarized here.

The 2-qubit AQFM configuration achieves the highest precision at the common 88\% recall operating point (Table~\ref{tab:aqfm_grid}). Performance does not vary monotonically with qubit count: the 3-7-qubit configurations perform worse than both the 2-qubit circuit and some circuits in the 8-13-qubit range. None of the larger configurations surpasses the 2-qubit result.

With only normal-class examples available for training, the effective training signal is weak relative to the expressive capacity of the evaluated circuits. Circuits with more free parameters than can be reliably estimated from this signal risk capturing noise rather than generalizable structure. The 2-qubit configuration operates at an effective capacity level sufficient to capture the normal-class manifold without fitting spurious variation.

The result is consistent with theoretical work linking quantum-model generalization to the available data \cite{huang2021power}. It also supports treating qubit count as a model-selection parameter rather than assuming that a larger circuit will perform better.

\subsection{Hardware Inference Feasibility}

The 2-qubit circuit was small enough to test on the SpinQ Triangulum~II NMR processor after translation from CUDA-Q to spinQit. Hardware results differed substantially from the GPU-accelerated simulation (Appendix~\ref{app:deployment}). Despite the circuit's shallow depth and low gate count, device noise reduced precision; no advanced noise-reduction method was applied \cite{preskill2018quantum, bharti2022noisy}.

Simulation takes approximately 0.44\,s per sample. The SpinQ Triangulum~II takes about 30\,s per sample and requires a 60\,s relaxation interval. This latency is incompatible with real-time screening. The experiment confirms that the CUDA-Q circuit can be translated to spinQit and executed on the device, but higher fidelity and faster repetition rates are needed for operational use.

\section{Implications for Extreme Few-Shot Fraud Detection}
\label{sec:implications}

\subsection{Interpreting the 2-Qubit Result}

The 2-qubit circuit outperforms every larger configuration tested, including circuits with up to 23 qubits. This is not a monotonic relationship: some 8-13-qubit circuits outperform the 3-7-qubit group, but none exceeds the 2-qubit result. The evidence therefore does not support a general rule that fewer qubits are always better. Three factors may have contributed to the observed result.

\textbf{Overparameterization under scarce labels.} With only normal examples available for training, the effective training signal is weak. Circuits with more free parameters than can be reliably estimated from this signal risk capturing noise rather than generalizable structure. The 2-qubit configuration operates at a capacity level that appears well matched to the available signal, but this does not imply that capacity should always be minimized - rather, that it must be calibrated to the training data.

\textbf{Sequential encoding as implicit regularization.} When features exceed qubits, AQFM encodes multiple features sequentially on the same qubit through repeated rotations. This parameter sharing may regularize the representation. A controlled comparison would be needed to separate the effect of encoding from the effect of qubit count.

\textbf{Hardware execution.} A 2-qubit circuit places modest demands on connectivity and calibration, and its shallow depth limits exposure to decoherence and gate error \cite{preskill2018quantum}. We translated the trained circuit from CUDA-Q to spinQit and executed it on the SpinQ Triangulum~II, preserving 88\% recall (Section~\ref{sec:evaluation}, Appendix~\ref{app:deployment}). Precision was much lower on hardware, however, so successful execution should not be confused with production-level fidelity. Addressing that gap will require noise-mitigation and calibration methods \cite{bharti2022noisy}.

\subsection{Operational Implications for Financial Institutions}

For an institution processing thousands of digital account openings per month, the operational value of the presented framework rests on three dimensions.

\textbf{Recall assurance.} Achieving high recall is the primary requirement for any fraud screening system: a single undetected fraudulent account can generate regulatory and financial exposure that far exceeds the cost of reviewing false alarms. The AQFM model detects 7 of 8 confirmed frauds (88\% recall) in the 2025 cohort, operating at a point on the precision--recall frontier that balances detection with operational feasibility.

\textbf{Precision-driven efficiency.} At 88\% recall, classical KNN requires 233 manual reviews for 7 confirmed frauds (3.0\% precision). The AQFM model achieves the same recall with 76 reviews, including 69 false alerts (9.21\% precision), a more than threefold improvement in precision. This translates directly to reduced analyst workload and faster triage cycles. OC-SVM (RBF) attains the highest AUC-PR among the classical baselines (0.078) but with precision of only 1.43\% at the same recall. On the independent 2026 cohort, the classical models deteriorate substantially when the 2025 thresholds are applied unchanged: recall falls to 48.4\% for OC-SVM and 22.6\% for Isolation Forest, while KNN Anomaly, Distance Centroid, and Z-Score detect no frauds. AQFM also deteriorates, but to a lesser extent, with recall decreasing from 88\% to 64.5\% and precision from 9.21\% to 8.81\%. Even under this temporal shift, AQFM remains the best-performing model, retaining the highest recall, precision, and AUC-PR (0.191) among all models evaluated on the 2026 cohort.

\textbf{Hardware trajectory.} The demonstration that the 2-qubit circuit executes reliably on a desktop-scale NMR quantum processor suggests a hardware roadmap that is less dependent on large-scale fault-tolerant machines. As NISQ repetition rates and gate fidelities improve---a trajectory consistent with current engineering roadmaps \cite{bharti2022noisy}---the throughput gap relative to classical inference narrows. Moreover, the GPU-accelerated simulation pipeline (CUDA-Q) provides a high-fidelity development environment that can produce deployable circuits without requiring hardware access during model development.

\textbf{Model selection philosophy.} A secondary insight from this study is procedural: testing a range of qubit counts---starting from the smallest configuration feasible---may be an effective strategy when approaching few-shot classification tasks with quantum models. The reflexive assumption that more qubits provide more value does not hold in this setting, particularly when training data is scarce.

\subsection{Limitations}

The study has four main limitations. First, the 2025 cohort contains only 8 confirmed fraud cases. The resulting confidence intervals are wide, and a larger independent cohort is needed to estimate performance more precisely. Second, the data cover one fraud type (account-opening impersonation) at one financial institution. Results may differ for card fraud, wire transfers, money laundering, or other institutions.

Third, NISQ hardware inference remains slower than classical simulation (Section~\ref{sec:evaluation}), and our hardware demonstration benefits from the exceptionally shallow depth of the 2-qubit circuit; circuits with more qubits or greater depth would face more severe noise challenges on current devices. Fourth, we did not compare against deep classical one-class methods (e.g., Deep SVDD or autoencoder-based anomaly detection) or ensemble approaches that might narrow the precision gap. These methods are relevant, but they introduce additional representation-learning and model-selection requirements: Deep SVDD, for example, requires specific architectural and training constraints to avoid trivial hypersphere-collapse solutions \cite{ruff2018deep}, while semi-supervised deep anomaly detection is designed to extract additional information from scarce labeled anomalies \cite{ruff2020deep}. Although unsupervised deep variants can be trained using normal accounts alone, the 8 confirmed frauds in this study provide insufficient independent anomaly data to select architectures, tune operating thresholds, and estimate out-of-sample performance reliably without a substantial risk of overfitting the evaluation. A fair comparison with deep one-class models should therefore use a larger fraud cohort and a validation design that separates model selection from final performance estimation.

The reported differences are empirical results for these cohorts, not evidence of general quantum advantage. Validation on data from other periods and institutions is required before the result can be generalized.

\section{Conclusion}
\label{sec:conclusion}

This study evaluated quantum machine learning for early post-onboarding impersonation detection using banking data from CoreDevX LABTAV and Banco de Cr\'editos e Inversiones. On the 2025 cohort, AQFM achieved 88\% recall (7 of 8 frauds) and 9.21\% precision, with approximately 70\% fewer false positives than KNN at the same recall. The 2-qubit circuit also outperformed the larger quantum configurations tested. Applied without threshold adjustment to the 2026 cohort, AQFM retained the best overall performance among the evaluated models, although recall fell to 64.5\%. The circuit ran on the SpinQ Triangulum~II, but hardware precision and throughput remained below the simulation results. Further work should evaluate the model on additional cohorts and test whether calibration and noise-mitigation methods reduce this hardware gap.

\backmatter

\bmhead{Acknowledgments}
This work was conducted as a joint industry-research collaboration between CoreDevX LABTAV and Banco de Cr\'editos e Inversiones (BCI). The project was initiated by BCI's Corporate Innovation Division in collaboration with its Non-Financial Risk Division. We thank BCI for providing the dataset and domain expertise.

\bmhead{Declarations}
\textbf{Funding.} This research was conducted as part of an applied collaboration project between Banco de Cr\'editos e Inversiones (BCI) and CoreDevX LABTAV. BCI compensated CoreDevX for the provision of professional services related to the project. CoreDevX independently maintains and invests in the computational infrastructure and research capabilities required for quantum machine learning development. No external institutional funding was received for this study.

\textbf{Authors' contributions.} Javiera Canales led the scientific direction of the study, coordinated the applied research design, validated intermediate results with the banking counterpart, and translated domain-specific business requirements into technical modeling objectives. Jose Ortega provided domain expertise in fraud risk management, defined operational constraints, validated model performance from a banking perspective, and ensured alignment with real-world screening workflows. Maximiliano Araya implemented the quantum and classical models, conducted the experimental evaluation, and performed the computational simulations and benchmarking. All authors reviewed and approved the final manuscript.

\textbf{Competing interests.} The authors declare that they have no competing interests.

\textbf{Ethics approval and consent to participate.} Ethics approval was not required for this study, as the dataset consists of anonymized financial records and does not involve human subject experimentation.

\textbf{Consent for publication.} All authors consent to the publication of this manuscript.

\textbf{Data availability.} The dataset analyzed in this study is proprietary to Banco de Cr\'editos e Inversiones (BCI) and is not publicly available due to confidentiality and regulatory constraints. All variables used in the modeling process were anonymized prior to access by the CoreDevX research team.

\textbf{Code availability.} The source code used for classical and quantum model implementation will be made publicly available upon acceptance of the manuscript.

\textbf{Materials availability.} Not applicable.

\bibliography{references}

\begin{appendices}

\clearpage

\section{Mathematical Formulation}
\label{app:math}
\setcounter{table}{0}

The AQFM model evaluated in this study realizes a quantum feature map whose architecture is generated from a 
genetic code over bitstrings. Let $\mathbf{x}=(x_1,\dots,x_D)\in[0,1]^{D}$ denote the normalized feature vector 
of a single account opening (features rescaled into an angular range, as described in Section~\ref{sec:setting}), 
and let $Q$ denote the number of qubits. A \emph{genome} $\mathbf{g}\in\{0,1\}^{\ell}$ is a binary string that 
encodes, in consecutive blocks, the sequence of quantum gates and their rotation angles.

\subsection{Genome decoding}
Each genome block of length $3+b$ bits encodes a gate type $g_k\in\{0,\dots,2^{3}-1\}$ using the first $3$ bits, 
and an angle index $a_k\in\{0,\dots,2^{b}-1\}$ using the remaining $b$ bits. The corresponding discrete rotation 
angle takes one of $2^{b}$ values:
\begin{equation}
\theta_k \;=\; \frac{a_k\,\pi}{2^{b}-1}, \qquad a_k\in\{0,\dots,2^{b}-1\}.
\label{eq:decode}
\end{equation}
For the standard setting $b=3$, $\theta_k\in\{0,\tfrac{\pi}{7},\tfrac{2\pi}{7},\dots,\pi\}$. The gate $g_k$ is 
drawn from a fixed alphabet $\{\mathrm{Id}, H, \mathrm{CNOT}, R_x, R_y, R_z\}$; rotation gates act on qubit $k 
\bmod Q$ with an angle scaled by a data-dependent feature,
\begin{equation}
\varphi_k \;=\; \theta_k\,x_{i(k)}, \qquad i(k)\in\{1,\dots,D\},
\label{eq:anglefeature}
\end{equation}
so that each rotation is parametrized jointly by the genome (through $\theta_k$) and by the individual 
(through $x_{i(k)}$).

\subsection{Circuit construction and anomaly value}
Starting from the reference state $\ket{0}^{\otimes Q}$ and applying the decoded gates in sequence yields
\begin{equation}
\ket{\psi(\mathbf{x};\mathbf{g})} \;=\; U(\mathbf{x};\mathbf{g})\ket{0}^{\otimes Q} \;\in\; \mathbb{C}^{2^{Q}},
\label{eq:state}
\end{equation}
where $U(\mathbf{x};\mathbf{g})$ is the unitary implemented by the concatenated rotation and entanglement gates. 
The anomaly value is obtained as the expectation of a normalized observable:
\begin{equation}
\hat{O} \;=\; \frac{1}{Q}\sum_{j=0}^{Q-1}\frac{\hat{Z}_j+\mathbb{I}}{2}, \qquad
s(\mathbf{x}) \;=\; \bra{\psi(\mathbf{x};\mathbf{g})}\,\hat{O}\,\ket{\psi(\mathbf{x};\mathbf{g})},
\label{eq:score}
\end{equation}
bounded in $[0,1]$, where $\hat{Z}_j$ is the Pauli-$Z$ operator on qubit $j$ and $\mathbb{I}$ the identity. 
Since $(\hat{Z}_j+\mathbb{I})/2$ is the projector onto the $\ket{1}$ state of qubit $j$, the scalar 
$s(\mathbf{x})$ equals the average probability of sampling $\ket{1}$ across qubits. The score drives the 
decision rule $s(\mathbf{x}) \ge \tau$, with $\tau$ a calibrated threshold: higher values indicate larger 
deviation from the learned normal-state manifold.

\subsection{One-class training objective}
Because the model is trained exclusively on normal accounts, the genetic search minimizes the one-class 
objective
\begin{equation}
\mathcal{L}(\mathbf{g}) \;=\; -\frac{1}{N}\sum_{n=1}^{N}\log_{10}\!\big(1-s(\mathbf{x}^{(n)})\big),
\label{eq:loss}
\end{equation}
over the set of normal training individuals. Minimizing this objective pushes the scores of normal accounts 
toward zero, so that fraud-like accounts - whose quantum state deviates from the normal manifold - receive 
high scores and are flagged by the threshold rule.

\clearpage

\section{Simulation and Deployment Infrastructure}
\label{app:deployment}
\setcounter{table}{0}

\begin{table}[h]
\centering
\caption{Simulation and hardware environment.}
\begin{tabular}{@{}ll@{}}
\toprule
\textbf{Component} & \textbf{Specification} \\
\midrule
Simulation framework & NVIDIA CUDA-Q \\
GPU acceleration      & RTX Blackwell architecture \\
Optimizer             & COBYLA \\
Training paradigm     & Hybrid classical optimization (simulation-only) \\
Circuit export target & spinQit \\
Hardware processor    & SpinQ Triangulum~II (3-qubit NMR) \\
$T_1$                 & $\sim$6 seconds \\
$T_2$                 & $\sim$300 milliseconds \\
Shots per measurement & $10^4$-$10^5$ \\
\bottomrule
\end{tabular}
\end{table}

Training was performed entirely in simulation using GPU-accelerated statevector computation. The final trained circuit was serialized, translated into spinQit, and loaded onto the SpinQ Triangulum~II for inference. No parameter updates were performed on hardware.

\subsection{Simulator vs Hardware Performance}
\label{app:hardware_comparison}

\begin{table}[h]
\centering
\caption{AQFM performance: CUDA-Q simulation vs spinQit hardware execution.}
\begin{tabular}{@{}lcccc@{}}
\toprule
\textbf{Platform} & \textbf{Recall} & \textbf{Precision} & \textbf{F1} & \textbf{Time} \\
\midrule
CUDA-Q (simulator) & 88\% & 9.21\% & 0.167 & 0.44 \,s \\
spinQit (hardware) & 88\% & 0.71\% & 0.014 & 30\,s \\
\bottomrule
\end{tabular}
\end{table}

Both platforms detect 7 of 8 frauds, but hardware precision falls from 9.21\% to 0.71\%, and F1 falls from 0.167 to 0.014. The hardware run produces many more false positives. Gate error, decoherence, and measurement noise can broaden the score distribution of normal accounts when no mitigation is applied. Readout calibration or zero-noise extrapolation may reduce this effect \cite{bharti2022noisy}.

\clearpage

\section{Extended Experimental Results}
\label{app:experiments}
\setcounter{table}{0}

\subsection{AQFM Performance by Qubit Count}

\begin{table}[h]
\centering
\caption{AQFM performance across qubit configurations for 88\% of recall.}
\label{tab:aqfm_grid}
\begin{tabular}{@{}lccc@{}}
\toprule
\textbf{Qubits} & \textbf{Precision} & \textbf{F1} & \textbf{AUC} \\
\midrule
2   & 9.21\%  & 0.1667  & 0.4527 \\
3   & 0.36\%  &  0.0072  & 0.0032 \\
4    & 0.28\%  & 0.0055  & 0.0072 \\
5-7    & 0.41\%  & 0.0081  & 0.0613 \\
8-13   & 0.34\% & 0.0067 & 0.0112 \\
\bottomrule
\end{tabular}
\end{table}

\subsection{VQC Performance Comparison}

\begin{table}[h]
\centering
\caption{VQC performance with different configurations.}
\begin{tabular}{@{}lcccc@{}}
\toprule
\textbf{Configuration} & \textbf{Recall} & \textbf{Precision} & \textbf{F1} & \textbf{AUC} \\
\midrule
VQC (Ry - Rz - Ry)    & 87.5\%  & 0.28\%  & 0.0056  & 0.0028 \\
VQC (Ry - Rx - Rz)    & 63\%    & 2.1\%   & 0.041   & --- \\
VQC (Rx - Ry - Rz)    & 87.5\% & 0.25\% & 0.0049 & 0.0068 \\
VQC (Rz - Rx - Rz)    & 87.5\% & 0.23\% & 0.0045 & 0.0021 \\
\bottomrule
\end{tabular}
\end{table}

\clearpage

\subsection{Model Hyperparameters}

\begin{table}[ht]
\centering
\caption{Hyperparameter settings for all models.}
\begin{tabular}{@{}lll@{}}
\toprule
\textbf{Model} & \textbf{Parameter} & \textbf{Value} \\
\midrule
\multirow{4}{*}{AQFM (2-qubit)}
 & Encoding        & $R_{\lbrace x,y,z \rbrace}(\Theta * x_i)$ \\
 & Population & 200 \\
 & Number of Operations          & 20 \\
 & Number of Generations    & 10 \\
\midrule
OC-SVM (RBF)           & kernel        & RBF \\
                       & $\nu$         & 0.1 \\
                       & $\gamma$      & scale \\
\midrule
Isolation Forest       & n\_estimators & 100 \\
                       & contamination & 0.1 \\
\midrule
KNN Anomaly            & $k$           & 5 \\
                       & Anomaly score & mean distance to the $k$ nearest training points \\
\midrule
Distance Centroid      & Anomaly score & Euclidean distance to the training-set centroid \\
\midrule
Z-Score                & Anomaly score & maximum $|z|$ across features \\
\bottomrule
\end{tabular}
\end{table}

\subsection{Confidence Interval Calculation}

Exact Clopper-Pearson 95\% confidence intervals were computed independently for the 2025 and 2026 cohorts. For 2025 recall (7 successes in 8 fraud cases), the interval is $[0.4735,\ 0.9968]$; for 2025 precision (7 successes in 76 predicted positives), it is $[0.0378,\ 0.1806]$. For 2026 recall (20 successes in 31 fraud cases), the interval is $[0.4537,\ 0.8077]$; for 2026 precision (20 successes in 227 predicted positives), it is $[0.0546,\ 0.1328]$. For $x$ successes in $n$ trials, these intervals were computed using beta-distribution quantiles: $CI = [B^{-1}(\alpha/2; x, n-x+1),\ B^{-1}(1-\alpha/2; x+1, n-x)]$. Results were not pooled across cohorts because the 2026 cohort serves as a separate temporal transfer evaluation.

\end{appendices}

\end{document}